\documentstyle[12pt,axodraw,subeqn]{article}
\def\neu {\tilde\chi_1^0}
\def\sl {\tilde{\ell}}

\def\msl {m_{\sl}}

\def\tanb {\tan \beta}
\def\beq{\begin{equation}}
\def\eeq{\end{equation}}

\def\gev{\;{\rm GeV}}
\newcommand{\lsim}
{{\;\raise0.3ex\hbox{$<$\kern-0.75em\raise-1.1ex\hbox{$\sim$}}\;}}
\newcommand{\gsim}
{{\;\raise0.3ex\hbox{$>$\kern-0.75em\raise-1.1ex\hbox{$\sim$}}\;}}
 
\begin{document}

\begin{flushright}
{HIP-2003-26/TH}\\
{MRI-P-030301}\\ 
{SHEP-03-04}\\
{\large \tt hep-ph/0304192}\\[3ex]
\end{flushright}
\begin{center}
{\Large\bf Slepton production from gauge boson fusion} \\[10mm]
{\bf Debajyoti Choudhury$^{(a)}$, 
Anindya Datta$^{(b)}$,
Katri Huitu$^{(b)}$,
Partha Konar$^{(a)}$,\\
Stefano Moretti$^{(c)}$} and 
{\bf Biswarup Mukhopadhyaya$^{(a)}$}  \\[2ex]

$^{(a)}${\em Harish-Chandra Research Institute,\\
Chhatnag Road, Jhusi, Allahabad - 211 019, India} \\[1ex]
$^{(b)}${\em Helsinki Institute of Physics, and\\
Division of High Energy Physics, Department of Physical Sciences\\
P.O. Box 64, 00014 University of Helsinki, Finland}\\[1ex]
$^{(c)}${\em Department of Physics \& Astronomy, University of Southampton,\\
Highfield, Southampton SO17 1BJ, UK}
\end{center}
\vskip.5in

\begin{abstract}

We emphasise that charged slepton pairs produced via vector-boson
fusion along with two high-mass, high-$p_T$ forward/backward jets (in
two opposite hemispheres) can have a higher production cross-section
for heavy slepton masses than that from conventional Drell-Yan
production at a hadronic collider like the LHC. We analyse the signal
and leading backgrounds in detail in the minimal supersymmetric
standard model with conserved baryon and lepton numbers. Our
investigation reveals that the mass reach of the vector-boson fusion
channel is certainly an improvement over the scope of the Drell-Yan
mode.

\end{abstract}

\section{Introduction}
Vector-boson fusion (VBF) at hadronic machines such as the Large
Hadron Collider (LHC) at CERN has been suggested as a useful channel
for studying Higgs boson signals. Characteristic features of
this mechanism are two highly energetic quark-jets, produced in the
forward/backward directions in opposite hemispheres of the detector and 
carrying
a large invariant mass.  The absence of colour exchange between these two
jets ensures a suppression of hadronic activity in the
central region \cite{bjorken}, contrary to the case of
typical QCD backgrounds. Though it was originally proposed as a
signal for a heavy Higgs boson \cite{dawson,abbasabadi}, the usefulness
of the VBF channel in detecting  an intermediate mass Higgs boson has also
been subsequently demonstrated \cite{zeppenfeld}.

Some recent works \cite{konar} have further pointed out 
the effectiveness of this channel in the context of new physics 
searches, particularly for new particles that do not interact strongly. 
Perhaps the best example is afforded by supersymmetric theories,
wherein conventional search strategies for neutralinos and charginos 
may run into difficulties, at least for a significant part of the parameter 
space. Encouraged by the success of the VBF channel
in exploring such cases, we investigate here its 
efficacy in the search for the supersymmetric partners of the
leptons, namely, the sleptons ($\tilde \ell$). 
This is of particular interest as the 
conventional search strategies for such particles at  the LHC are not 
very promising, especially for slepton masses above
300 GeV or so. In fact, direct pair production of  
sleptons via the Drell-Yan (DY)process has been investigated 
extensively in the literature~\cite{tata_slep, reno}.
At the Tevatron and the LHC, the corresponding next-to-leading order
(NLO) production cross-sections fall below 1 fb
for slepton masses above 200 and 500 GeV, respectively~\cite{reno}. 
Within, e.g.,
the minimal supergravity (mSUGRA) scenario, such sleptons would decay
mainly into a charged lepton ($\ell$) and the lightest neutralino 
($\neu$), thus
resulting in an opposite-sign di-lepton pair with missing 
energy\footnote{Slepton pair-production
  has also been investigated in the case of gauge mediated Supersymmetry
(SUSY) breaking
  \cite{mele} where the third generation of sleptons has a
  very distinct decay signature.}. 
As can be expected, 
the upper limit of the corresponding mass reach is quite low 
($\sim$ 250 GeV at the LHC) \cite{tata_slep}.

In the present work, we want to investigate whether the above mass
limit can be improved at the LHC when we use the VBF channel for
slepton production. Slepton pair-production via $WW$
fusion has been discussed earlier in ref. \cite{gunion}, and more
recently, in the special context of anomaly mediated SUSY breaking,  
in \cite{huitu}.  It is
needless to mention that the VBF channel is suppressed by four powers
of $g_{\rm{EW}}$, the electro-weak (EW) coupling, with respect to the
DY mode. However, for the latter, the cross-section falls rather fast
with the slepton mass, whereas one should expect a milder dependence
on $\msl$ in VBF.  We will only consider pair-production of charged
sleptons ($\tilde
\ell_L$, $\tilde \ell_R$) 
along with two forward/backward jets.  In most of the
following analysis, we will assume the general minimal supersymmetric
standard model (MSSM) with parameters defined at the EW scale, as
we will not adhere to any particular SUSY-breaking
scenario and make no assumption related to any high mass scale physics
other than adopting gauge coupling unification. This implies
that whereas the slepton masses are free parameters in our analysis,
the neutralino masses and couplings are completely specified by the 
$SU(2)$ gaugino mass $M_2$, the Higgs(ino) mass parameter $\mu$ and $\tan
\beta$, the ratio of the two Higgs vacuum expectation values
arising in the MSSM. 
The only constraints on this set of parameters are the 
experimental ones, most notably those imposed by the LEP analyses.
At the very end, we
will try to correlate our results to the mSUGRA parameter space.

The plan of the article is as follows. In section 2, we will
discuss in detail the nature of the proposed signal and its various
features. Section 3 will be devoted to a discussion of the event selection
criteria adopted here, while the end results, embodied in a set of
discovery contours, are reported in section 4. We summarise and conclude
in section 5.

\section{The Signal}
     \label{sec:signal}
\subsection{Slepton pair-production}
     \label{subsec:prodn}
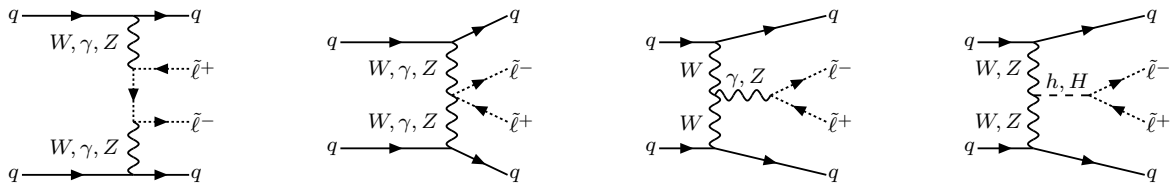
\begin{figure}[h] 
{
\unitlength=1.0 pt
\SetScale{1.0}
\SetWidth{0.7}      
\scriptsize    
\noindent
\begin{picture}(95,99)(0,0)
\Text(15.0,90.0)[r]{$q$}
\ArrowLine(16.0,90.0)(58.0,90.0)
\Text(80.0,90.0)[l]{$q$}
\ArrowLine(58.0,90.0)(79.0,90.0)
\Text(53.0,80.0)[r]{$W,\gamma,Z$}
\Photon(58.0,90.0)(58.0,70.0){2}{3}
\Text(80.0,70.0)[l]{$\tilde{\ell}^+$}
\DashArrowLine(79.0,70.0)(58.0,70.0){1.0}
\Text(54.0,60.0)[r]{$$}
\DashArrowLine(58.0,70.0)(58.0,50.0){1.0}
\Text(80.0,50.0)[l]{$\tilde{\ell}^-$}
\DashArrowLine(58.0,50.0)(79.0,50.0){1.0}
\Text(53.0,40.0)[r]{$W,\gamma,Z$}
\Photon(58.0,50.0)(58.0,30.0){2}{3}
\Text(15.0,30.0)[r]{$q$}
\ArrowLine(16.0,30.0)(58.0,30.0)
\Text(80.0,30.0)[l]{$q$}
\ArrowLine(58.0,30.0)(79.0,30.0)
\end{picture} \ 
{} \qquad\allowbreak
\begin{picture}(95,99)(0,0)
\Text(15.0,80.0)[r]{$q$}
\ArrowLine(16.0,80.0)(58.0,80.0)
\Text(80.0,90.0)[l]{$q$}
\ArrowLine(58.0,80.0)(79.0,90.0)
\Text(53.0,70.0)[r]{$W,\gamma,Z$}
\Photon(58.0,80.0)(58.0,60.0){2}{3}
\Text(80.0,70.0)[l]{$\tilde{\ell}^-$}
\DashArrowLine(58.0,60.0)(79.0,70.0){1.0}
\Text(80.0,50.0)[l]{$\tilde{\ell}^+$}
\DashArrowLine(79.0,50.0)(58.0,60.0){1.0}
\Text(53.0,50.0)[r]{$W,\gamma,Z$}
\Photon(58.0,60.0)(58.0,40.0){2}{3}
\Text(15.0,40.0)[r]{$q$}
\ArrowLine(16.0,40.0)(58.0,40.0)
\Text(80.0,30.0)[l]{$q$}
\ArrowLine(58.0,40.0)(79.0,30.0)
\end{picture} \
{} \qquad\allowbreak
\begin{picture}(95,99)(0,0)
\Text(15.0,80.0)[r]{$q$}
\ArrowLine(16.0,80.0)(37.0,80.0)
\Text(80.0,90.0)[l]{$q$}
\ArrowLine(37.0,80.0)(79.0,90.0)
\Text(33.0,70.0)[r]{$W$}
\Photon(37.0,80.0)(37.0,60.0){2}{3}
\Text(49.0,65.0)[b]{$\gamma,Z$}
\Photon(37.0,60.0)(58.0,60.0){2}{3}
\Text(80.0,70.0)[l]{$\tilde{\ell}^-$}
\DashArrowLine(58.0,60.0)(79.0,70.0){1.0}
\Text(80.0,50.0)[l]{$\tilde{\ell}^+$}
\DashArrowLine(79.0,50.0)(58.0,60.0){1.0}
\Text(33.0,50.0)[r]{$W$}
\Photon(37.0,60.0)(37.0,40.0){2}{3}
\Text(15.0,40.0)[r]{$q$}
\ArrowLine(16.0,40.0)(37.0,40.0)
\Text(80.0,30.0)[l]{$q$}
\ArrowLine(37.0,40.0)(79.0,30.0)
\end{picture} \ 
{} \qquad\allowbreak
\begin{picture}(95,99)(0,0)
\Text(15.0,80.0)[r]{$q$}
\ArrowLine(16.0,80.0)(37.0,80.0)
\Text(80.0,90.0)[l]{$q$}
\ArrowLine(37.0,80.0)(79.0,90.0)
\Text(33.0,70.0)[r]{$W,Z$}
\Photon(37.0,80.0)(37.0,60.0){2}{3}
\Text(50.0,63.0)[b]{$h,H$}
\DashLine(37.0,60.0)(58.0,60.0){3.0}
\Text(80.0,70.0)[l]{$\tilde{\ell}^-$}
\DashArrowLine(58.0,60.0)(79.0,70.0){1.0}
\Text(80.0,50.0)[l]{$\tilde{\ell}^+$}
\DashArrowLine(79.0,50.0)(58.0,60.0){1.0}
\Text(33.0,50.0)[r]{$W,Z$}
\Photon(37.0,60.0)(37.0,40.0){2}{3}
\Text(15.0,40.0)[r]{$q$}
\ArrowLine(16.0,40.0)(37.0,40.0)
\ArrowLine(37.0,40.0)(79.0,30.0)
\Text(80.0,30.0)[l]{$q$}
\end{picture} \ 
}
\caption{\em Generic parton level diagrams leading to slepton
  pair-production through electroweak VBF at hadronic colliders.
}
\label{fig:feyn}
\end{figure}

We begin by considering slepton pair-production through VBF. 
The generic (lowest-order) 
diagrams contributing to this process are depicted in 
Fig. \ref{fig:feyn}, where each of the $q$'s represents either a quark or an 
antiquark. Clearly, such diagrams do not exhaust the entire set of 
contributions to the process $q_1 q_2 \to 
q_3 q_4 \tilde \ell \tilde \ell^*$. In fact,
apart from a host of other EW diagrams, one also has to include
those involving a gluon exchange. In addition, although they
do not interfere with the signal, one also has to consider 
graphs with gluons in either of the initial and final state. 
 Although we shall
    impose kinematic constraints to ensure that diagrams such as those in
    Fig. \ref{fig:feyn} dominate overwhelmingly, in the actual
    computation,  one still needs to
    include the full set of diagrams that lead to a slepton-pair
    accompanied by two jets.
In doing so, we limit ourselves to a tree-level calculation and 
use the {\sc HELAS} subroutines~\cite{helas} to numerically evaluate the
ensuing helicity amplitudes. For our parton-level Monte Carlo analysis, 
we use the CTEQ4L parton distributions \cite{cteq} 
with the scale 
set at the slepton mass ($\msl$).

The very structure of the VBF diagrams immediately suggests that 
such contributions would be largely concentrated in kinematic 
regions where the vector-bosons are nearly on mass shell. 
This translates into two rather 
forward/backward jets, one in each hemisphere. 
Since no coloured particle is exchanged, the rapidity 
gap between these forward/backward jets would be essentially free of hadronic 
activity.  Thus we start by characterising the signal in terms of the following
basic criteria:

\begin{itemize}
\item[{\em (a)}] 
The sleptons (and their decay products) 
are entirely contained in the rapidity regime in between the 
two forward/backward 
jets, labeled as $j_i$ ($i=1$, 2), satisfying the following requirements: 
\subequations
\beq
2 \le |\eta(j_i)| \le 5 \ , \quad \eta(j_1)\; \eta(j_2) < 0 \ ;
        \label{cut_jetrap}
\eeq
\item[{\em (b)}] Both jets should have sufficient transverse momentum to be 
detected, namely,
\beq
p_{T}(j_i) \ge 15 \gev \ ;
        \label{cut_jetpt}
\eeq
\item[{\em (c)}] 
The invariant mass of the pair of forward jets should be sufficiently large,
\beq
M_{j_1j_2} > 650 \gev \ ;
        \label{cut_jetmass}
\eeq
\endsubequations
\item[{\em (d)}] 
There should be no hadronic activity in the rapidity interval between 
these two jets.
   \label{no_had_activ}
\end{itemize}
We have explicitly checked that, on imposition of the above criteria, 
the resulting cross-section is overwhelmingly dominated by the VBF 
diagrams. It should further be remembered that these are only our 
`basic cuts', and  serve the purpose of establishing
the characteristics of a VBF event. However, as we shall see shortly,
additional cuts are required to enhance the visibility of the signal
against backgrounds.
While, in our analysis, these criteria have been imposed at the parton 
level, they are expected to mimic actual detector events even after 
hadronisation is incorporated. Experimentally, criterion {\em (d)} 
is implemented by applying  a central jet veto.
The above cuts select signal events whose survival probability against 
such a veto turns out to be
between 80 to 90 percent \cite{eboli}. For {\em real emission correction
to DY-type processes} (which involve colour exchange between the jets),
in contrast, the corresponding survival probability is below 
30 per cent \cite{rainwater}. In the remainder of our analysis, 
we will include the full set of contributions, weighed 
appropriately by the respective survival probabilities.

\begin{figure}[t]
\centerline{
\epsfxsize= 8 cm\epsfysize=8.0cm
                     \epsfbox{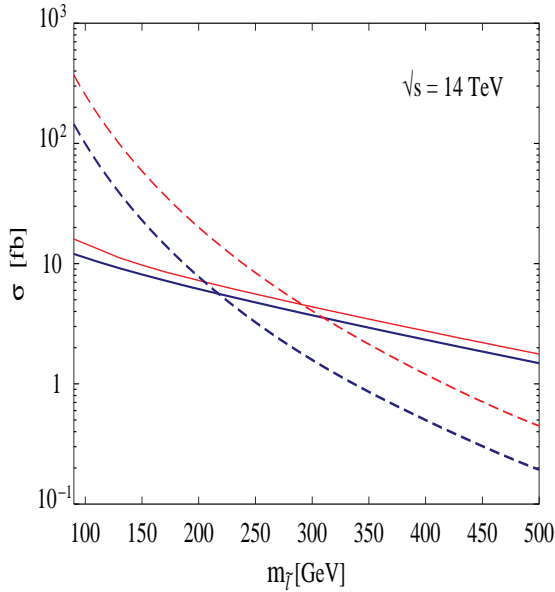}
}
\caption{\em The total cross section (solid lines) for slepton
        pair-production at the LHC in association with two forward jets. 
        The cuts of eqns.(\protect\ref{cut_jetrap}--\protect\ref{cut_jetmass}) 
        have been imposed on the VBF rates. 
        The CTEQ4L parton distributions have been used
        with the factorization scale set at $\msl$. The dashed curves 
        represent the corresponding DY cross sections. In each case,
        the upper and lower curves correspond to $\sl_L$ and $\sl_R$ 
        (one flavour) respectively.
}
        \label{fig:tot_cs}
\end{figure}

Let us now examine the total production cross-section and the possible 
parameter dependences of the signal process. 
Since we shall concentrate only on the sleptons 
of the first two generations ($\tilde e_{L, R}$ and $\tilde \mu_{L, R}$), 
the Higgs mediated diagrams in Fig.~\ref{fig:feyn} are not important. 
This also implies that the 
production cross-section is not sensitive to either $\mu$,  
$\tan \beta$ or the slepton 
mixing\footnote{However, as we shall see later, $\mu$ plays a 
        significant role in slepton decays and affects then the 
        signal as a whole.}. 
Thus the production cross-section is essentially model-independent and 
is determined solely by the slepton mass $\msl$. In Fig. \ref{fig:tot_cs},
we display this functional dependence. The lowest order 
DY cross-section (without any cuts) is also shown for an approximate
comparison of the relative magnitude. A few points are 
immediately obvious.

\begin{itemize}
\item Formally, our cross-section is suppressed by 
        two powers of $\alpha_{\rm em}$($\alpha_{\rm weak}$) when 
	compared to the DY one. 
	This is reflected in the dominance of the DY rates  
	for small slepton masses.
\item The cross-section fall-off with mass is much slower for the VBF 
        process, as compared to the DY mode, as intimated.  This can
        be understood by recognising that the DY cross-section suffers
        from the presence of an $s$-channel propagator. In contrast,
        the VBF process could be viewed in terms of an effective
        $\gamma/Z/W$ approximation, wherein the large logarithms
        associated with the emission of a ``nearly massless'' gauge
        boson compensate for the extra factors of $\alpha_{\rm
        em}$($\alpha_{\rm weak}$).Notice however that such logarithmic
        enhancements are finite and well under control (that is, they
        need not a higher order treatment) since the requirements of
        forward/backward jet tagging that we will put in place (a
        minimum $p_T$ together with a maximal rapidity) act as
        effective regulators, on the same footing as in
        Refs. \cite{dawson,abbasabadi,zeppenfeld}.

\item The VBF process is dominated by the photon diagrams. This is to be 
        expected in view of the previous remark and is reflected by the 
        relatively small ($\le 10\%$) 
        fractional difference in the cross-sections for 
        $\sl_L$ and $\sl_R$ production.
\end{itemize}
As Fig. \ref{fig:tot_cs} also shows,
the VBF cross-section is significantly larger than the 
DY one for large values of $\msl$. Since this is precisely the 
region of the parameter space where the DY production mode is of little use, 
it behoves us to investigate the VBF channel further. In addition, the 
two forward/backward jets are peculiar to this channel and could serve to 
eliminate backgrounds.

\subsection{Slepton decay modes and kinematics}
     \label{subsec:decay}
Once produced, the sleptons will decay into either a chargino-neutrino
pair or into a neu\-tra\-li\-no-lepton pair. The partial decay widths are 
governed by both the mass and composition of the charginos (neutralinos) 
as well as the handedness of the slepton ($L$ or $R$). As is well known, 
as long as $R$-parity is conserved, the lightest
supersymmetric particle (LSP) is stable. Since consistency with 
observations demands that the lightest neutralino be the LSP, 
the latter is invisible
and all other supersymmetric particles decay into it. Thus the slepton decay
must result in {\em same-flavour opposite-sign di-lepton pairs 
associated with 
missing transverse momentum}. Cascade decays through the heavier 
neutralinos/charginos would produce a similar signature (with still
more particle tracks in the detector), so that they may  
be deemed as part of the signal. 
However, for reasons explained later, we will primarily be 
concentrating on the direct decay of the slepton into the lepton-LSP pair. 
As we have mentioned in Sec.~\ref{subsec:prodn}, we would be requiring 
the {\it lepton pair to lie within the rapidity interval between the jets}. 
In other words ($i=1,2$),
\subequations
\beq
  \left| \eta(\ell_i) \right| \le 2 \ .
         \label{cut_leprap}
\eeq
Of course, the two leptons must have enough transverse momenta to be 
detectable:
\beq
    p_T(\ell_i) \geq 15 \gev \ .
        \label{cut_leppt}
\eeq
\endsubequations

\begin{figure}[!h]
\centerline{
\epsfxsize= 9 cm\epsfysize=8.0cm
                     \epsfbox{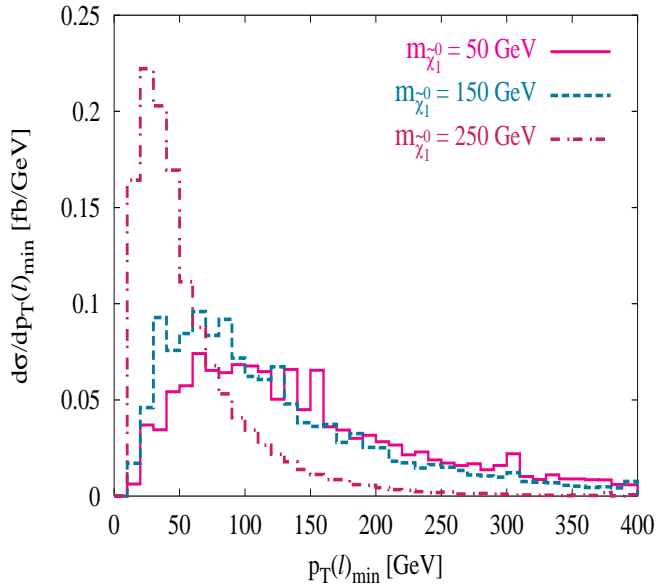}
}
\caption{\em The distribution of the softer of the two lepton $p_T$'s 
for three different values of the LSP mass and a fixed slepton mass of 
300 GeV.}
   \label{fig:lepton_pt}
\end{figure}
Before we decide on  further selection criteria, 
it is useful to examine the signal profile resulting from the 
production of a slepton pair of a given mass and decaying into a 
particular neutralino, again of a given mass. A variable of interest 
here is the smaller of the two lepton transverse momenta, namely, 
$ {\rm{min}}[p_T(\ell_1), p_T(\ell_2)]$. In Fig. \ref{fig:lepton_pt}, 
we display the distribution in this observable for a given slepton 
mass and three
representative values of the LSP mass. Note that a smaller value of 
$\msl - m_{\neu}$ softens this distribution. This is to be expected.
Since the sleptons prefer to be produced with little transverse momenta, 
a high $p_T$ for the decay products would only be possible if the 
mass difference were large. A similar pattern would appear 
in the case of the missing transverse momentum. 

This also explains partly our `neglect' of cascade decays through the 
heavier neutralinos and charginos. The final states resulting from such 
decay channels typically contain additional leptons or jets. 
To avoid QCD backgrounds, we would need to concentrate on the multi-lepton
modes. Such decay patterns, however, occur less frequently than those 
involving quarks (and hence additional jets). Moreover, with a smaller 
mass difference between the slepton and a heavier neutralino/chargino, 
the primary lepton would tend to be softer and hence often evade the 
selection process. Explicit computation shows that the inclusion of the 
cascade decays can result in only a marginal improvement of our results
and we shall ignore their effects henceforth. 

\begin{figure}[!t]
\vspace*{4ex}
\centerline{
\epsfxsize= 16 cm\epsfysize=12.0cm
                     \epsfbox{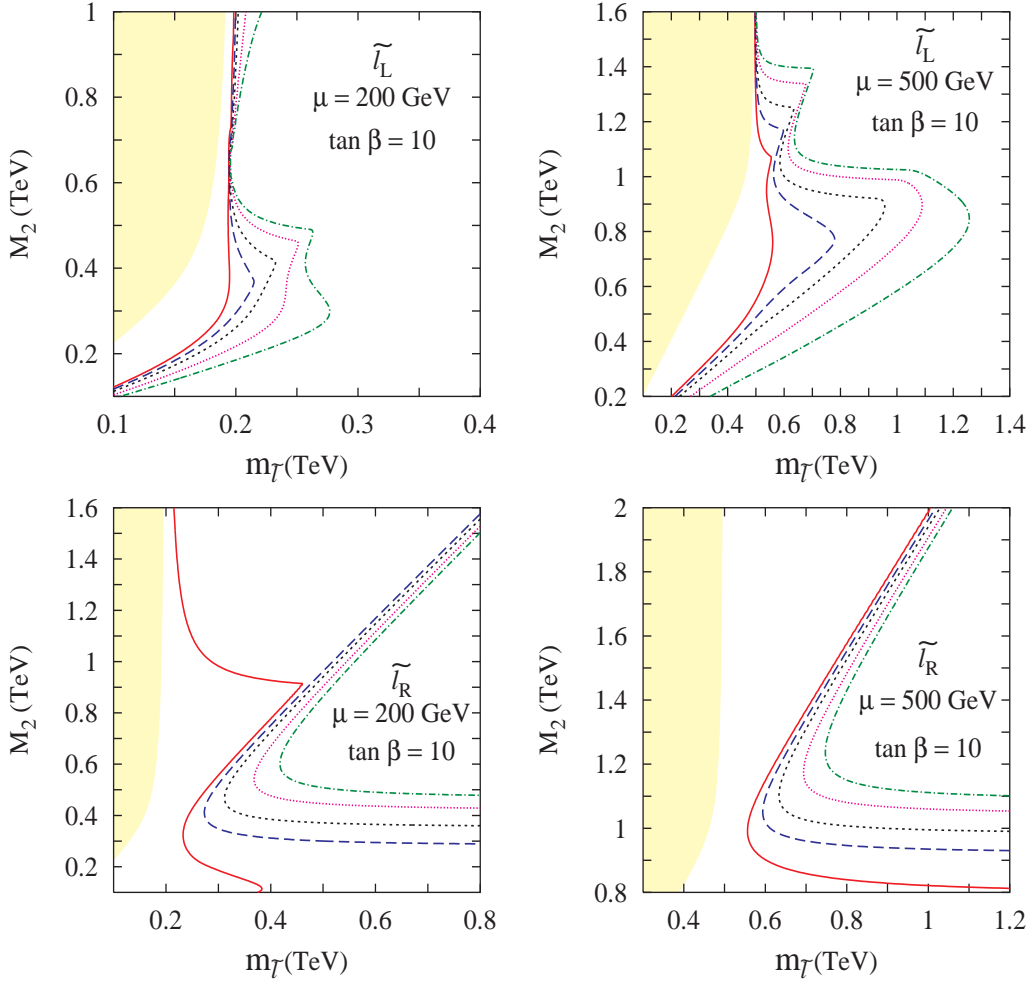}
}
\caption{\em Contours for constant $BR (\sl \to \neu + \ell)$ for 
both $\sl_L$ (upper panels) and $\sl_R$ (lower panels) in the 
$\msl$--$M_2$ plane. In each case, the left and right panels correspond 
to $\mu = 200 \gev$ and $500 \gev$, respectively. 
The shaded area corresponds to the part of the parameter space that 
leads to the slepton being the LSP. 
The contours, from left to right, are for $BR (\sl \to \neu + \ell)
 = 0.9, 0.7, 0.5, 0.3$ and   $0.2$, respectively.}
\label{fig:decay}
\end{figure}

\subsection{Slepton Branching Fractions}
    \label{subsec:brfrac}

We now turn to the issue of the slepton branching ratio (BR) into the lightest 
neutralino. This depends on quite a few parameters: $\msl$, $\mu$, $\tan \beta$
and the gaugino mass parameters $M_1$ and $M_2$. Of these, the dependence 
on $\tan \beta$ is the least pronounced and therefore we shall 
henceforth use only one value of it, namely, $10$. 
Furthermore, to reduce the number of parameters, we shall assume the 
unification relation between $M_1$ and $M_2$. Thus, only three parameters 
remain, namely $\msl$, $\mu$ and $M_1$. 
For a given slepton mass, the relevant branching fraction is then governed 
by essentially two factors: ($i$) the composition of the LSP and 
($ii$) whether decays into the heavier neutralinos/charginos are allowed. 
The resulting dependence is still quite intricate as can be gauged from 
Fig. \ref{fig:decay}, where we present iso-branching fraction contours 
in the $\msl$--$M_2$ plane for two positive values of $\mu$. 
A set of conclusions follow immediately. 
\begin{itemize}

\item For a given mass, the $\sl_R$ has a larger probability for decaying
directly into the
LSP as compared to the $\sl_L$. This effect is even more pronounced for 
larger $\mu$ and can be understood from the fact that whereas 
the $\sl_R$ has no coupling to the $\widetilde W^{\pm, 0}$ eigenstates, 
it is precisely these states that the $\sl_L$ preferentially decays into. 

\item For $\mu < M_2$, the 
two lightest neutralinos and the lighter chargino are often Higgsino-dominated.
Selectrons and smuons then tend to cascade through the heavier neutralinos
(heavier chargino). However, this possibility is curtailed when $M_2$
is large so that kinematic accessibility of these states is denied.

\item When $\mu$ and $M_2$ are comparable, the relative weight of the Bino
and Higgsino states in the LSP controls the BR of sleptons
decaying into it.

\end{itemize}

As we have already seen (Fig. \ref{fig:tot_cs}), the production cross-sections
for $\sl_L$ and $\sl_R$ are very similar, with the former being slightly 
larger. However, with the $\sl_R$ decaying into the LSP much oftener, 
it is expected that, for identical masses, it is this ($\sl_R \sl_R^*$)
production channel that will finally dominate the signal.

\subsection{Signal profile and parameter dependence}
    \label{subsec:profile}
\begin{figure}[t]
\centerline{
\epsfxsize= 8 cm\epsfysize=8.0cm
                     \epsfbox{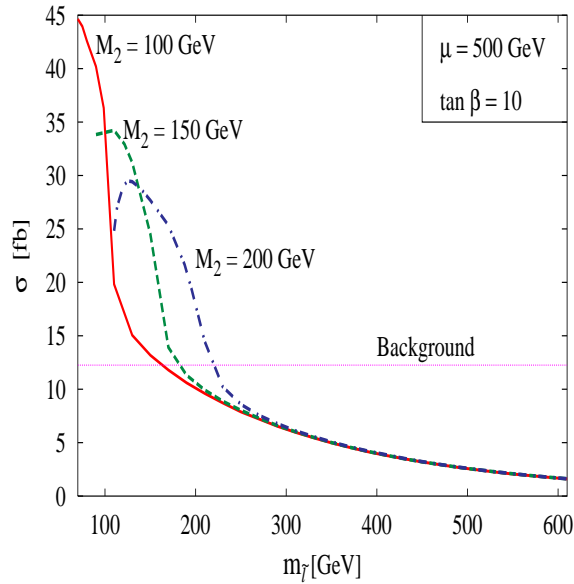}
}
\caption{\em Variation of the signal cross-section with the 
slepton mass for some representative values of $M_2$ and given 
$\mu$ and $\tan \beta$. The corresponding 
LSP masses are 
 $m_{\neu} = 48.8, \ 73.6, \ 98.4 \gev$ respectively.
Only the  basic cuts of 
eqns.(\protect\ref{cut_jetrap}--\protect\ref{cut_jetmass},
\protect\ref{cut_leprap}--\protect\ref{cut_leppt}) have been imposed.
Also shown is the corresponding background cross-section.
}
\label{fig:m2_var_of_cs}
\end{figure}

Before we end this section, we would like to discuss the interplay of 
the kinematic effects between the slepton-LSP mass difference 
 (as exemplified by Fig. \ref{fig:lepton_pt})
and the branching fractions. In doing this we shall assume that the two 
sleptons $\sl_{L,R}$ are degenerate, a very good approximation in 
SUGRA-inspired scenarios. In Fig. \ref{fig:m2_var_of_cs}, we demonstrate
the dependence of the cross-section on the slepton mass for three
representative values of $M_2$. For $\msl \gg M_2 $, the $\sl_R$ 
decays predominantly into the LSP while the $\sl_L$ is allowed
more channels. The important point, however, is that, in this 
limit, the branching fraction into the LSP is essentially independent of 
$M_2$. Moreover, with a large separation between $\msl$ and $m_{\neu}$,
the leptons acquire transverse momenta sufficiently large  
(Fig. \ref{fig:lepton_pt}) to satisfy the selection criteria. Thus, in 
this regime, the cross-section is practically independent of $M_2$ and is 
determined solely by $\msl$. For very low values of $\msl$, on the other hand, 
the aforementioned kinematic dependence on the mass difference becomes 
very important: the larger $M_2$ is, the smaller is the average value of 
$p_T(\ell)$, resulting in the suppression of the signal 
(Fig. \ref{fig:m2_var_of_cs}). And finally, the very sharp decrease 
in the signal strength for $\msl \gsim M_2$ can be traced to the 
rapid change of the branching fraction into the LSP on account 
of new channels opening up.

\section{Backgrounds and their elimination}

Same-flavour, opposite-sign di-lepton and missing energy signals at the LHC
can be faked by standard model (SM) processes where two opposite-sign
$W$'s or $\tau$'s are produced with two forward/backward jets, with the $W$'s
 or
$\tau$'s decaying leptonically.  There is also a source of reducible
background from $ZZ$ production in presence of initial state radiation.  
Here, however, an invariant mass cut on the lepton pair can remove the 
latter background almost completely.
The continuum production due to an off-shell $Z$ going to leptons is
too small to be of any consequence.
Production of $t \bar t$ pairs with subsequent semi-leptonic decays of
top quarks can also produce the di-lepton + jets + missing transverse 
energy final state. We can easily get rid of this background though, by 
remembering that the jet associated with  top decay is always a $b$-jet. 
Such backgrounds are appreciable only for
$|\eta_j| \le 3$. Thus they can be eliminated with
a $b$-veto if the $b$-trigger works up to such a rapidity.
The pair-production of charged Higgs bosons in VBF
\cite{moretti} can also yield opposite-sign di-leptons with
missing transverse momentum and forward/backward jet activity.
This noise may be particularly dangerous, as it has the same topology
of the signal, including the reduced hadronic activity in the
central region. However, electrons and muons can emerge from charged
Higgs boson decays only indirectly via $\tau$'s, and hence with a leptonic 
BR suppression and in flavour combinations of equal probability. 
In the end, we have explicitly checked, by
varying the Higgs mass and the other relevant supersymmetric parameters 
consistently with the signal, that this 
background is not very large in general, 
so that we need not consider it any further.  
In summary, the dominant
contributions to the background come from $W$'s and
(direct) $\tau$'s in almost equal
strength, although some sizable effect is unavoidable from real
emission corrections to the DY process, despite its moderate
central jet veto survival probability.
We have estimated all these backgrounds using the package
{\sc MADGRAPH} \cite{madgraph}. With the cuts described above, the missing
transverse momentum and opposite sign di-electron/di-muon total background
comes out to be about 13 fb (choosing the factorisation scale at $2\,M_W$),
which we represent by the horizontal line in Fig. 5.  
Assuming that the b-veto will work upto $\eta~=~3$, the
       background gets contributions on the order of 8.5 fb and 4.5 fb
from $\tau\tau$ and $WW$, respectively. In the
following we will see that the background level can be reduced
significantly with a minimal sacrifice of the signal by exploiting suitable
kinematic distributions.

\begin{figure}[h]
\centerline{
\epsfxsize= 8,5 cm\epsfysize=7.0cm
                     \epsfbox{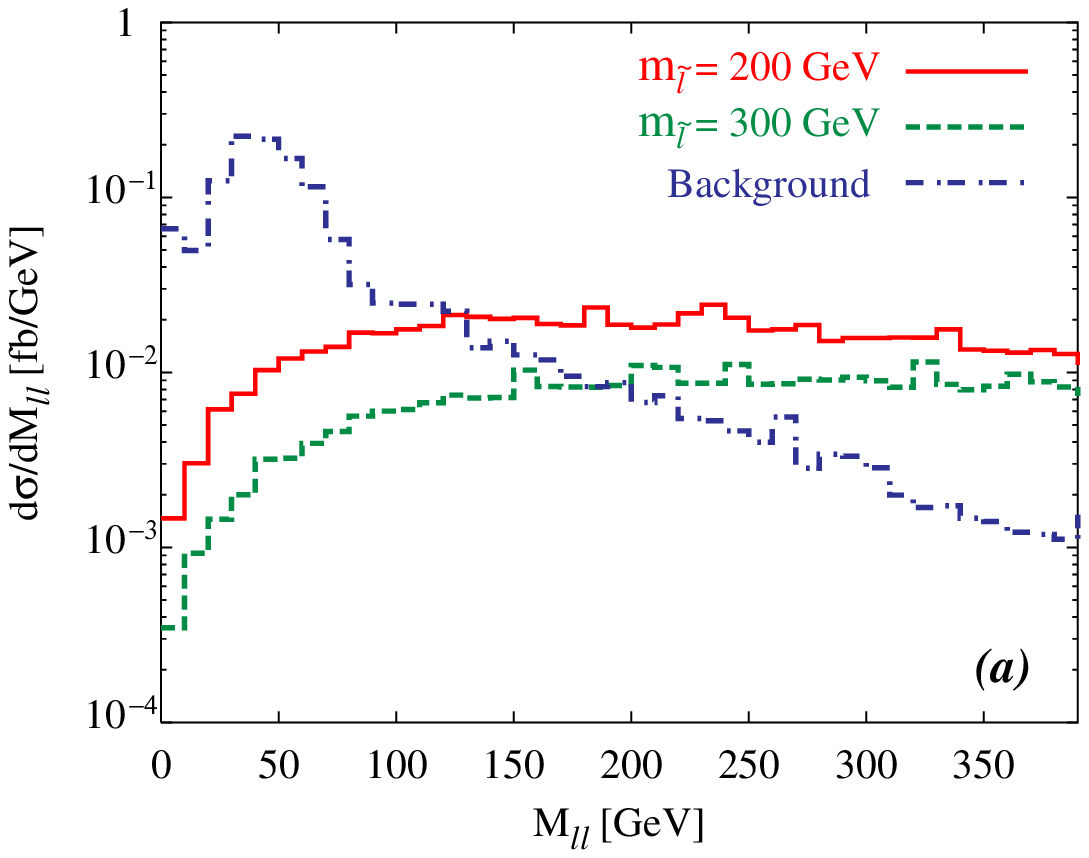}
        \hspace*{.7cm}
\epsfxsize=8.5 cm\epsfysize=7.0cm
                     \epsfbox{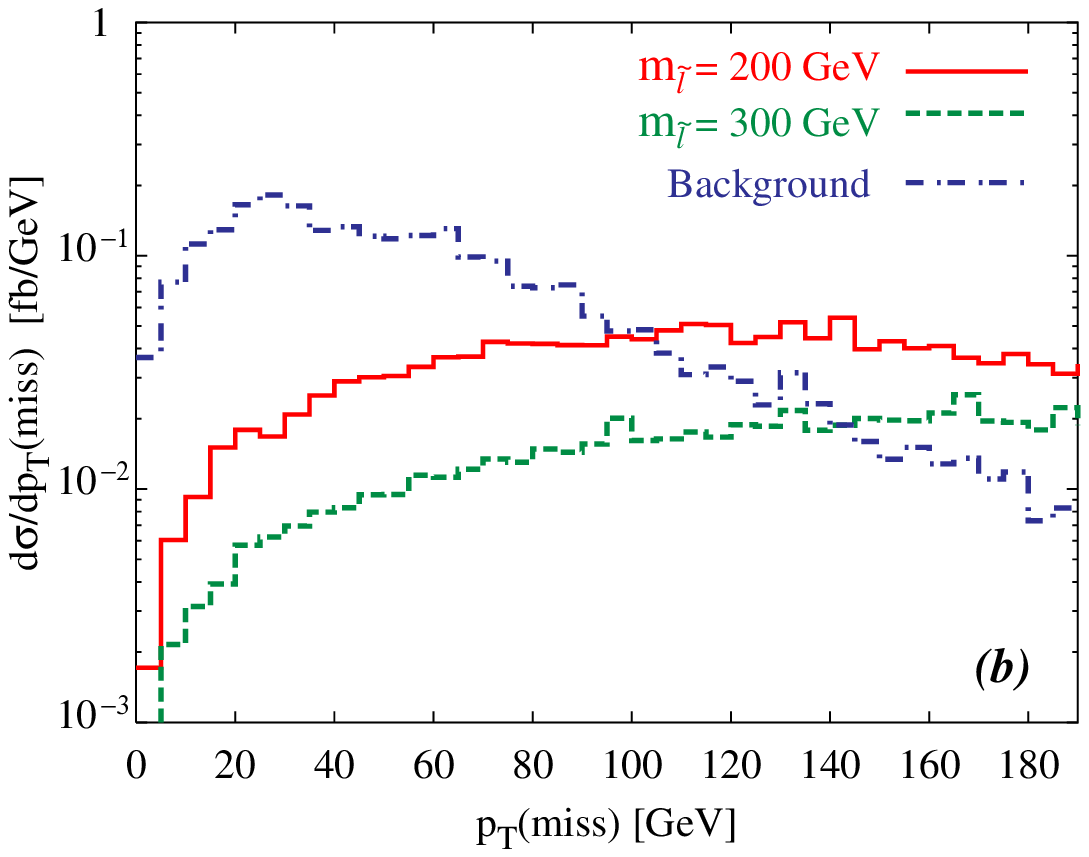}
}

\caption{{\it (a) Invariant mass $M_{\ell\ell}$ and (b) 
missing transverse momentum ${p}_T(\rm{miss})$ distributions for 
the signal for $M_2=150$ {\rm GeV}, $\mu= 500$ {\rm GeV}, 
$\tanb=10$, $\msl=200,300$ {\rm GeV}. Also shown are the corresponding 
background distributions.}}
\label{fig:et_mll}
\end{figure}
%
We examine the spectra in invariant mass
(Fig. \ref{fig:et_mll}a) of the di-lepton pair as well as in missing 
transverse energy
 (Fig. \ref{fig:et_mll}b) for both the signal (for some illustrative
values of $m_{\tilde \ell}$ and $M_{\mathrm{LSP}}$) and the combined
backgrounds after the previously mentioned cuts, and   observe that

\begin{itemize}

\item The invariant mass of the di-lepton pair has a much harder 
spectrum for the signal as compared to the total background. 

\item The missing transverse momentum
 distribution is harder for the signal, with the 
peaks shifting to higher values for lower masses of the LSP, once the
slepton mass is fixed. 

\end{itemize}
 
Keeping all this in mind, we impose a few additional 
selection criteria. For one, an event must be accompanied by a 
substantial missing transverse momentum:
\subequations
\beq
   p_T({\rm{miss}}) \geq 50 \gev \ .
    \label{cut_missET}
\eeq
Furthermore, while the invariant mass for the di-lepton pair should be 
sufficiently large to remove most of the background, i.e.  
\beq
       M_{\ell \ell} \geq 60 \gev \ ,
   \label{cut_mll}
\eeq
it should nevertheless be well away from the $Z$-mass (in order 
to eliminate backgrounds accruing from $p p \to j j Z \nu_i \bar{\nu_i}$):
\beq
     \left| M_{\ell \ell} - M_Z \right| > 5 \Gamma_Z \ .
    \label{cut_Zpole}
\eeq
\endsubequations

\begin{figure}[h]
\centerline{
\epsfxsize= 8.5 cm\epsfysize=7.0cm
                     \epsfbox{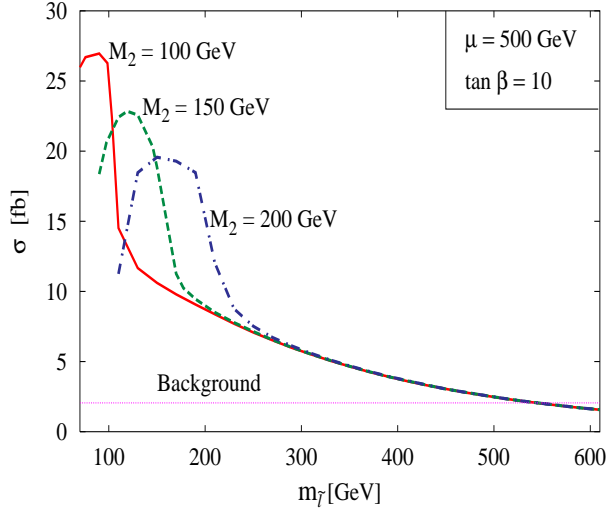}
}

\caption{\it As in Fig.~\protect\ref{fig:m2_var_of_cs}, but now the 
cuts of eqns.(\ref{cut_missET}--\ref{cut_Zpole}) have been imposed as well.}
\label{fig4}
\end{figure}

These extra cuts have only a moderate effect on the signal while
reducing the background down to only $\sim 2$ fb, as is evident from 
Fig. \ref{fig4}. As for the signal, the effects of the new
kinematic cuts are more pronounced for low mass sleptons. 
If we increase the neutralino
mass, the missing energy spectrum becomes harder while the di-lepton mass
distribution becomes softer. One can see by comparing Figs. 5 and 
7 that, for $\msl \sim$ 100 -- 200 GeV,
such a trade off has affected the $M_2 =$ 100 GeV case most severely.

It should also be mentioned here that the characteristic signals of
sleptons studied by us are subject to vitiation by other SUSY
processes, such as cascades from squarks, gluinos and electroweak
gauginos, leading to a potential `residual SUSY background'.  As
has already been noted in the first reference of \cite{konar}, the 
squark/gluino background can
be suppressed by the invariant mass cut on the forward jet pair.
Furthermore, a veto against central hadronic activities is also
helpful in eventually suppressing fake signals from squarks and
gluinos.  As for electroweak gauginos, in general their contributions
to the signals under scrutiny have been found to be smaller
\cite{konar}, mostly due to suppression by the leptonic branching
ratios of gauginos\footnote{A probable caveat is offered by a spectrum
      wherein the squarks are very heavy, while sleptons are only
      somewhat heavier than gauginos (with $\mu$ being relatively large).},
and the requirement
that both leptons in the final state be of the same flavour. One
situation where gauginos can intervene is when they can decay into
{\em real sleptons}. Such a case, however, again leads to
characteristic signals of the sleptons themselves, and therefore our
estimate, if anything, is of a conservative nature.

As has already been mentioned, one has to multiply the signal rates
with the central jet veto survival probability. This is a source of
theoretical uncertainty in the predictions; we have used as our
guidelines the results given in reference [14] for the survival
probabilities for electroweak and QCD processes, already noted in section
2.1. These probabilities pertain to a central jet with a minimum $p_T$ of
20 $GeV$, which therefore translates into a definition of hadronic
activities in the central region. For further discussion on the subject, 
the reader is directed to reference \cite{khoze}.

\section{Discovery contours}

\begin{figure}[h]
\centerline{
\epsfxsize= 8.5 cm\epsfysize=7.0cm
                     \epsfbox{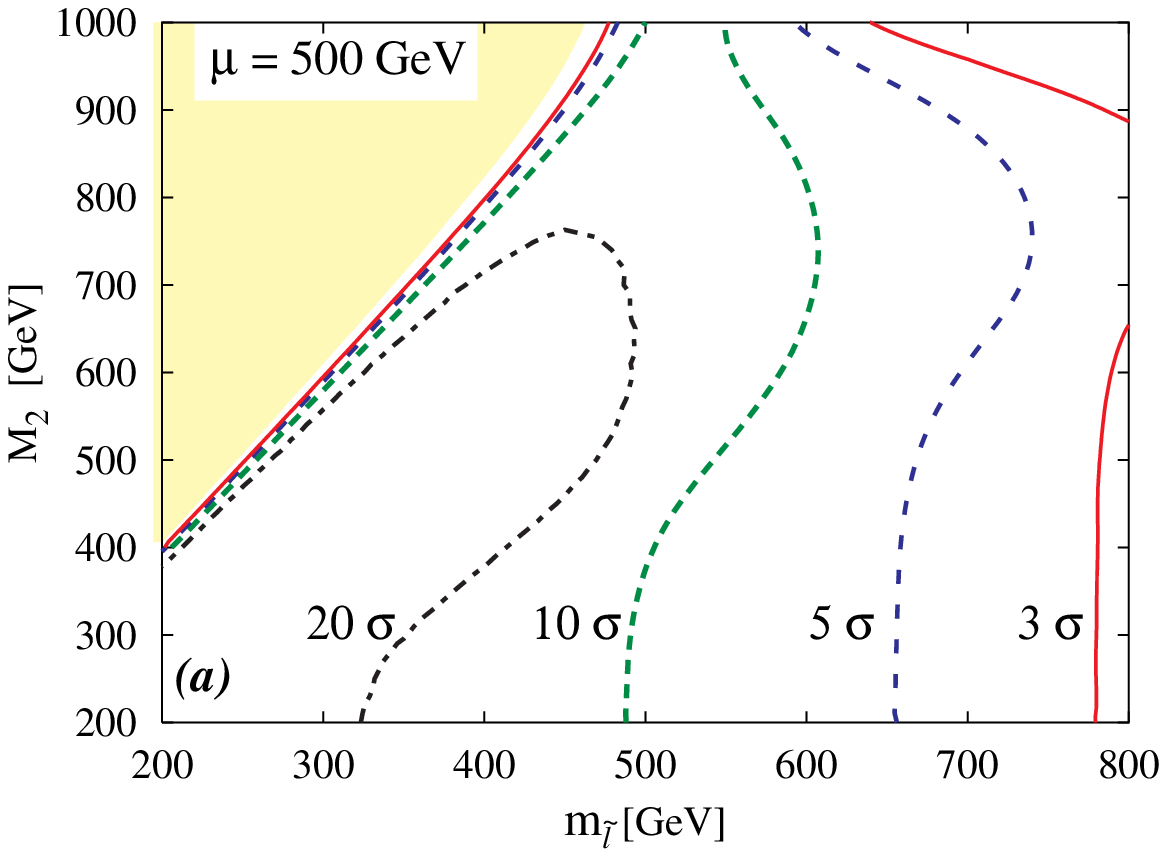}
        \hspace*{.3cm}
\epsfxsize=8.5 cm\epsfysize=7.0cm
                     \epsfbox{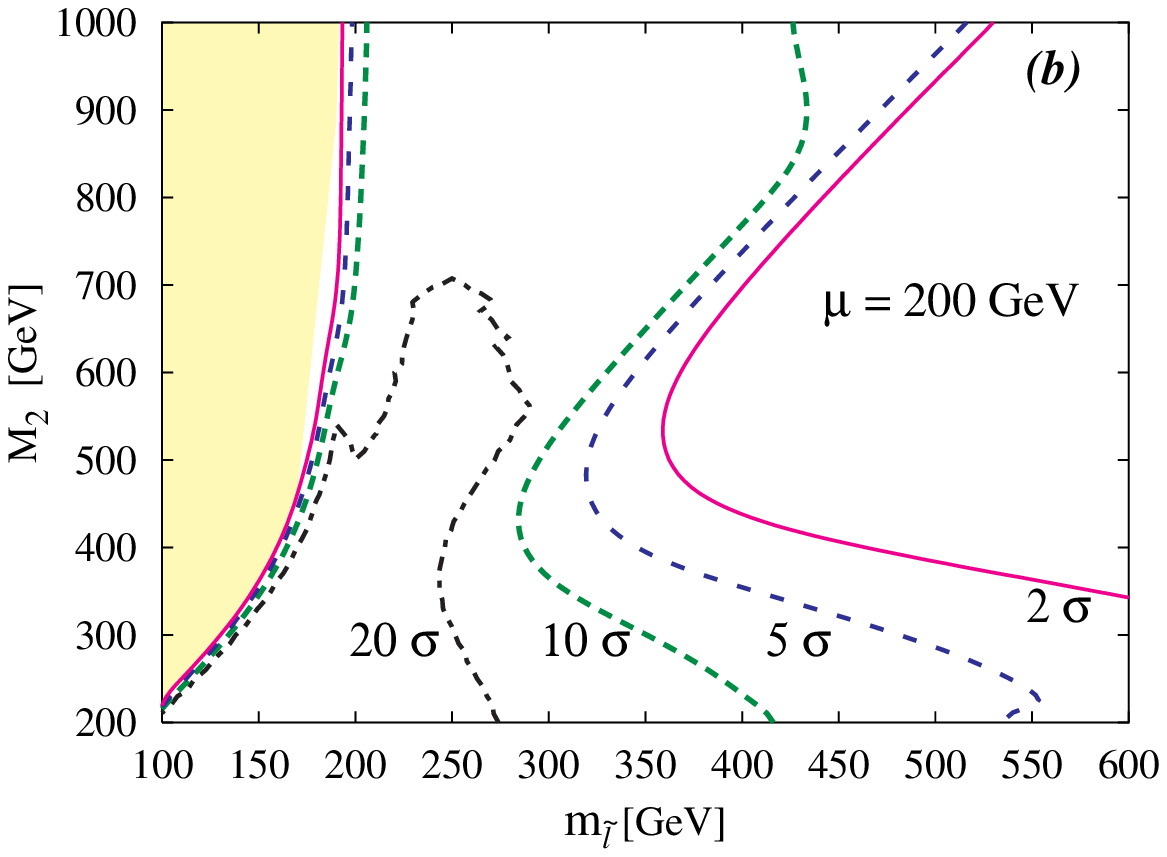}
}

\caption{{\it Contours of constant significance in the $M_2 - \msl$ plane for 
(a) $\mu = 500$ {\rm GeV} and (b) $\mu = 200$ {\rm GeV}, with $\tanb=10$.   
  }}
\label{fig:contours}
\end{figure}

We are now in a
position to predict the potential of our channel to explore/exclude
the supersymmetric parameters involved in this analysis. In 
Fig.\ref{fig:contours}, 
we present some significance contours of the predicted signals 
in the $M_2 - \msl$ plane for two values of $\mu$. The shaded regions
in the contour plots are either disallowed by the LEP data or 
inconsistent with the hypothesis that the lightest neutralino is the LSP. 
To calculate the significance
($\equiv S/\sqrt{B}$) we have assumed an integrated luminosity of
30 fb$^{-1}$. With 2.05 fb of total background cross-section, this 
implies 40 signal events for 5$\sigma$ discovery.

Evidently, the contours reflect rather promising statistics over a large
region of the parameter space.  The detailed nature of the contours
are mostly governed by features related to slepton production and decay,
which have been discussed in the previous sections. While there is a
complex interplay of different factors, we would like to recall at this
stage a few salient points which have roles to play in the predictions:

\begin{itemize}
\item The slepton production rates decrease with increasing slepton mass.

\item The composition of the LSP as well as the other neutralinos
and charginos is a deciding factor.

\item The mass difference between the slepton and the LSP determines
the hardness of the resulting leptons and therefore the survival probability
of the events against cuts.

\item As has been discussed earlier, while the right 
sleptons decay overwhelmingly into
a Bino-dominated LSP, the left ones often tend to cascade through 
the $SU(2)$ coupling.

\item The characteristic turning around of the curves for $\mu~=~200$ GeV
can also be seen for $\mu~=~500$ GeV for higher values of $M_2$ and $\msl$. 
\end{itemize}
\begin{figure}[h]
\centerline{
\epsfxsize= 9 cm\epsfysize=8.0cm
                     \epsfbox{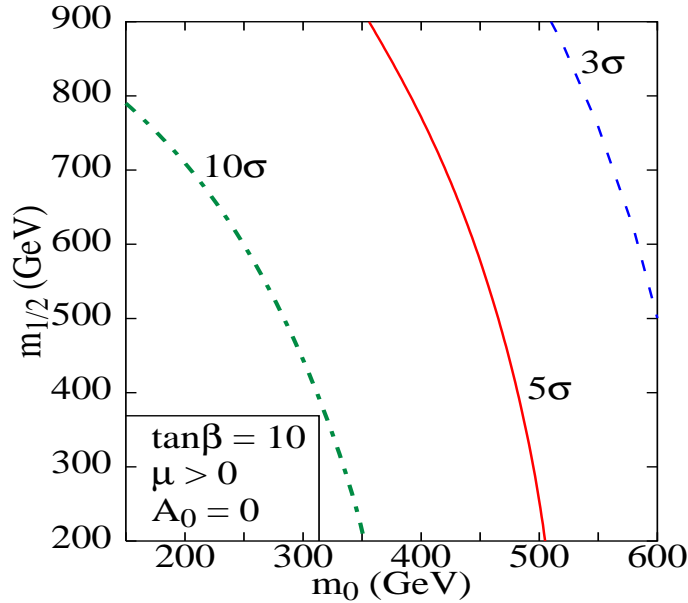}
}

\caption{\it Contours of constant significance in the
$M_{1/2} - m_0$ plane for $\mu > 0$, $A_0 = 0$ and $\tan \beta = 10$.  
An integrated luminosity of 30 fb$^{-1}$ has been assumed.}
\label{fig:sugra}
\end{figure}

The study of this signal also allows one to
draw significance contours in the parameter space of an
mSUGRA theory. For the purpose of illustration we have chosen $\mu >
0$, $A_0 = 0$ (always with $\tan \beta = 10$), where $A_0$ is the trilinear 
SUSY-breaking parameter at the unification scale. In Fig. \ref{fig:sugra}, the
significance contours are presented for three values of $S/\sqrt{B}$.  We
do not present the results for $\mu < 0$; it has already been stressed
that the sign of $\mu$ has very little effect on either the slepton 
pair-production cross-section or the
slepton decay BR to the LSP. The effects of 
slepton mixing or $\tan \beta$ are also negligible, since we are considering  
sleptons of the first two generations. 
We also assume radiative EW symmetry breaking.

The dissimilarity between the contours of Fig.~\ref{fig:sugra}
and  Figs.~\ref{fig:contours} might seem puzzling at first. However, 
an analytical study of the parameter space dependence immediately reveals 
the cause. Increasing $m_0$ (common scalar mass at the unification
scale), results in an increase in the values of $\msl$ and $\mu$.
The consequent (modest) enhancement of the branching ratio into the
LSP is, however, more than offset by the decrease in the production
cross-section due to higher slepton mass, and by the opening of
additional decay channels into higher neutralinos/charginos, so long
as $m_{1/2}$ is on the lower side.  It should be noted that such
channels (such as those into the second lightest neutralino and the
lighter chargino) affect the left sleptons more in the form of reduced
signal rates.  Increasing $m_{1/2}$ (the common gaugino mass at the
unification scale), on the other hand, has a twofold effect. First, it
increases the LSP mass thus affecting the decay kinematics. More
importantly, it also increases the slepton masses, preferentially that
of the left-sleptons. Since the latter suffer $SU(2)$ interactions
(unlike their right-handed counterparts), their mass increases with
$m_{1/2}$ at a faster rate. As a consequence, within this framework,
the production rate for the left sleptons falls faster with an
increasing $m_{1/2}$ than is the case for the
right-sleptons\cite{martin}.  Finally, for most of the parameter space
depicted in Fig.~\ref{fig:sugra}, the branching ratio for the
right-sleptons into the LSP is nearly unity, whereas the corresponding
one for the left-sleptons is a rather sensitive function of $(m_0 -
m_{1/2})$.  Together, these two factors result in the signal strength
being dominated by the contribution from the right-sleptons.
Moreover, with the decay kinematics playing a relatively subservient
role, the signal is determined largely by the mass of the right-handed
slepton alone. Thus the contours in Fig.~\ref{fig:sugra} largely
reflect the behaviour of right sleptons, particularly when $m_0$ and
$m_{1/2}$ are on the higher side.

For the kind of signal we are proposing, it is very crucial to know
the background normalisation very accurately, as one has to decide
about discovery/exclusion on the basis of counting the number of
events.  It is worthwhile to mention that, as we have only the leading
order (LO)
cross-section for the background, there is quite a strong dependence
of the latter upon the choice of the scale of $\alpha_{\rm s}$ and
also of the factorisation scale. However, should the
actual background normalisation 
be calculated directly from the LHC data and, without going into further
detail, one can legitimately assume a $5 \%$ uncertainty in our estimate 
of the background, by 
adding this error in quadrature to the estimated fluctuation of the 
latter, the requirement of 40 signal events for a 5$\sigma$
discovery would go up to only  42 events, which hardly implies any
modification to the mass reach and the event contours outlined above.

Before we conclude, let us compare our results with those in
ref. \cite{tata_slep}. The authors in \cite{tata_slep} calculated the
DY slepton pair-production and decay to have a di-lepton + missing
energy signal in the final state.  As already emphasised,
the slepton production
cross-section via the DY channel is more than
an order of magnitude higher than that 
via VBF for low slepton masses. However, the signal strength in the DY 
channel  falls rapidly as the slepton mass increases, and ultimately the
number of events becomes smaller than in the  VBF channel.  This is 
clearly evident from the slepton mass reach at the LHC ($\simeq 250$~GeV)
obtained in \cite{tata_slep}, whereas we have shown that the VBF channel can
easily probe slepton masses well up to 500 GeV with more than 5$\sigma$
significance over the leading backgrounds.

\section{Conclusion}
To summarise, we have investigated slepton pair-production via
VBF at the LHC.  The cross-section
for slepton pair-production along with two forward/backward jets has been
estimated at the parton level. For low mass
sleptons, the cross-section in the VBF channel is much
smaller than
the one from DY production of sleptons. However, for higher slepton
masses, the latter falls off quickly (below 1 fb) and the former becomes
dominant while remaining sizable. The pair-production cross-sections
for both left- and right-sleptons have then been estimated, the former being
marginally greater than the latter over the whole slepton mass range we
have considered.  We have then concentrated on slepton decays
to the lightest neutralino, leading to two unlike-sign
di-leptons (of same flavour) + missing transverse momentum along with two
forward/backward jets in the final state. Finally, we 
have devised simple kinematic cuts minimising the leading SM backgrounds
and found a rather large discovery potential up to
slepton masses on the order of $500$ GeV.
Although our analysis  was primarily based on the general MSSM, 
one can easily relate our results to the parameters of the
mSUGRA scenario, as we have done ourselves in one instance.  The overall 
conclusion is that our proposed signal should
help in increasing the slepton mass reach at the LHC in a significant 
manner, in comparison to the scope of the previously considered 
DY channel.

\section*{Acknowledgment} 

We thank the participants and organisers of the
7th Workshop on High Energy Physics Phenomenology (WHEPP-7) held
at Allahabad, India, where  this project was initiated.
DC thanks the Deptt. of Science \& Technology, India
for financial assistance under the Swarnajayanti Fellowship grant.
AD and KH thank the Academy
of Finland (project number 48787) for financial support. The work of 
BM is partially supported by the Board of Research in Nuclear Sciences, 
Government of India. 

\newcommand{\plb}[3]{{Phys. Lett.} {\bf B#1}, #2 (#3)}                  %
\newcommand{\prl}[3]{Phys. Rev. Lett. {\bf #1}, #2 (#3) }        %
\newcommand{\rmp}[3]{Rev. Mod.  Phys. {\bf #1}, #2 (#3)}             %
\newcommand{\prep}[3]{Phys. Rep. {\bf #1}, #2 (#3)}                   %
\newcommand{\rpp}[3]{Rep. Prog. Phys. {\bf #1}, #2 (#3)}             %
\newcommand{\prd}[3]{Phys. Rev. {\bf D#1}, #2 (#3)}                    %
\newcommand{\np}[3]{Nucl. Phys. {\bf B#1}, #2 (#3)}                     %
\newcommand{\npbps}[3]{Nucl. Phys. B (Proc. Suppl.)
           {\bf #1}, #2 (#3)}                                           %
\newcommand{\sci}[3]{Science {\bf #1}, #2 (#3)}                 %
\newcommand{\zp}[3]{Z.~Phys. C{\bf#1}, #2 (#3)}  
\newcommand{\epj}[3]{Eur. Phys. J. {\bf C#1}, #2 (#3)} 
\newcommand{\mpla}[3]{Mod. Phys. Lett. {\bf A#1}, #2 (#3)}             %
 \newcommand{\apj}[3]{ Astrophys. J.\/ {\bf #1}, #2 (#3)}       %
\newcommand{\jhep}[2]{{Jour. High Energy Phys.\/} {\bf #1} (#2) }%
\newcommand{\jpg}[3]{{J. Phys.\/} {\bf G#1}, #2 (#3)}%
\newcommand{\astropp}[3]{Astropart. Phys. {\bf #1}, #2 (#3)}            %
\newcommand{\ib}[3]{{ibid.\/} {\bf #1}, #2 (#3)}                    %
\newcommand{\nat}[3]{Nature (London) {\bf #1}, #2 (#3)}         %
 \newcommand{\app}[3]{{ Acta Phys. Polon.   B\/}{\bf #1}, #2 (#3)}%
\newcommand{\nuovocim}[3]{Nuovo Cim. {\bf C#1}, #2 (#3)}         %
\newcommand{\yadfiz}[4]{Yad. Fiz. {\bf #1}, #2 (#3);             %
Sov. J. Nucl.  Phys. {\bf #1} #3 (#4)]}               %
\newcommand{\jetp}[6]{{Zh. Eksp. Teor. Fiz.\/} {\bf #1} (#3), #2;
           {JETP } {\bf #4} (#6) #5}%
\newcommand{\philt}[3]{Phil. Trans. Roy. Soc. London A {\bf #1}, #2
        (#3)}                                                          %
\newcommand{\hepph}[1]{hep--ph/#1}           %
\newcommand{\hepex}[1]{hep--ex/#1}           %
\newcommand{\astro}[1]{(astro--ph/#1)}         %

\end{document}